\documentstyle[12pt,lathuile,epsfig]{article}
\newcommand{\be}{\begin{equation}}
\newcommand{\ee}{\end{equation}}
\def\lnf{{\bf  Up~Through }}
\def\le{{\bf  In~Up }}
\def\boa{{\bf  Up~Stop }}
\def\bob{{\bf  In~Down }}
\def\bo{{\bf  Up~Stop~+~In~Down }}
\def\atti{{\it attico }}
\def\tof{{\it T.o.F. }}
\def\mutau{ $\nu_\mu \leftrightarrow\nu_\tau$ }
\def\15{\raise.9ex\hbox{1-5)\kern-0.47em}\kern.4em}
\def\69{\raise.9ex\hbox{6-9)\kern-0.47em}\kern.4em}

\begin{document}

\title{MEASUREMENT OF THE ATMOSPHERIC MUON NEUTRINOS\\
WITH THE MACRO DETECTOR}
\author{Paolo Bernardini for the MACRO Collaboration\\
{\em Dipartimento di Fisica dell'Universit\`a and INFN}\\
{\em via per Arnesano, 73100 Lecce, Italy}\\}
\maketitle
\baselineskip=14.5pt
\begin{abstract}
The flux of muons induced by atmospheric neutrinos has been measured 
with the MACRO detector. Different event topologies have been detected, 
due to neutrino interactions in the apparatus and in the rock below it. 
The upward-throughgoing muon sample is the larger one and is generated 
by neutrinos with a peak energy of $\sim~100\ GeV$. The observed 
upward-throughgoing muons are $26\ \%$ fewer than expected and the 
zenith angle distribution does not fit with the expected one. Assuming 
\mutau neutrino oscillation, the angular shape and the normalization 
factor suggest maximal mixing and $\Delta m^2$ of a few times 
$10^{-3}\ eV^2$. Also the other event categories induced by low-energy 
neutrinos (peak energy $\sim 4\ GeV$) show a deficit of observed events. 
The value of this deficit and its uniformity with respect to the angular 
bins are in agreement with the oscillation parameters suggested by the 
analysis of the upward-throughgoing muon sample. 
\end{abstract}
\baselineskip=17pt

\vspace{3.3cm}

\hspace{3cm}
{\bf Les Rencontres de Physique de la Vallee d'Aoste}

\hspace{3cm}
{\bf February 28 - March 6, 1999, La Thuile, Italy}

\newpage
\section{Introduction}
\begin{figure}[bh]
 \begin{center}
  \vspace{-1.5cm}
  \mbox{\epsfig{file=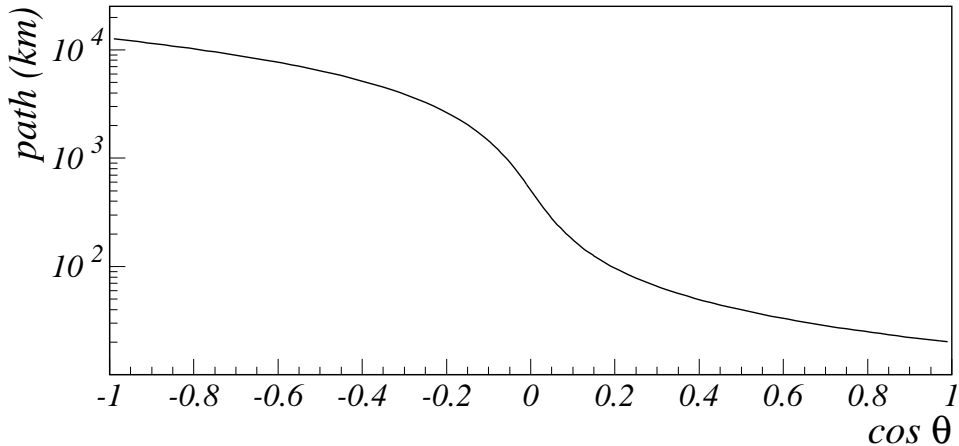,width=14cm}}
  \vspace{-7.0cm}
  \caption{\em Path of atmospheric neutrinos versus $\cos~\theta$. The 
               upward going $\nu$'s (negative values of $\cos~\theta$) 
               travel for thousands of kilometres before reaching an
               underground detector. 
  {\label{fig:path}}}
  \vspace{-0.5cm}
 \end{center}
\end{figure}

The interest in precise measurements of the flux of neutrinos
produced in cosmic ray cascades in the atmosphere has been growing
over the last years due to the anomaly in the ratio of contained
muon neutrino to electron neutrino events\15.
%% \cite{hira}\cite{casp}\cite{bec}\cite{all}\cite{fuku}
The anomaly finds explanation in the scenario of neutrino oscillation. 
The effects of neutrino oscillations appear also at 
higher energies, as reported by many experiments\69.
%% \cite{ah95}\cite{amb2}\cite{hata}\cite{fuku2}. 
The flux of muon neutrinos in the energy region from a few $GeV$ 
up to hundreds of $GeV$ is inferred from measurements of upgoing
muons. The flux of upgoing muons is reduced as a
consequence of $\nu$ oscillations. %% but the measurement of the total
%% number is not very crucial because of the high uncertainty. 
A clearer signature of $\nu$ oscillations in the range 
$10^{-3} < \Delta m^2 < 10^{-2}\ eV^2$ is connected with the
dependence of the reduction on the polar angle $\theta$ with
respect to the zenith. The reduction in the number of events is 
stronger near the vertical than near the horizontal directions
due to the longer pathlength of neutrinos from production to observation 
near the nadir (Fig.~\ref{fig:path}). 

Furthermore the flux of atmospheric muon neutrinos in the region
of a few $GeV$ can be studied looking at muons produced inside the
detector and muons externally produced and stopping inside it. If
the atmospheric neutrino anomalies are the results of neutrino
oscillations, it is expected a reduction in the flux of upward-going 
low-energy atmospheric neutrinos of about a factor of two, but 
without any distortion in the shape of the angular distribution. 

In this paper the MACRO flux measurements are updated for high and 
low energy muon neutrinos. The new results are in agreement with 
the previous ones\cite{ah95}\cite{amb2}, and the indication for
the $\nu$ oscillation hypothesis is now stronger.

\section{The MACRO detector}
The MACRO detector\cite{ah93} is located in the Gran Sasso
Laboratory, with a minimum rock overburden of $2700\ hg/cm^2$. It
is a large rectangular box ($76.6 \times 12 \times 9.3\ m^3$)
divided longitudinally in $6$ supermodules and vertically in a
lower and an upper part, called {\it attico}. The active elements
(see Fig.~\ref{fig:topo}) are liquid scintillation counters for
time measurement and streamer chambers for tracking, with $27^\circ$
stereo strip readouts. In the lower half of the detector streamer 
tube planes are alternated with trays of crushed rock absorbers.
The \atti is hollow and it is used as a work area in which
the electronics racks are placed. The streamer tube system allows
to achieve a tracking resolution for muons in the range 
$0.2^\circ$~--~$1^\circ$ as a function of the track length. Hence
the angular uncertainty is lower than the angular spread due to multiple 
scattering in the rock for a muon coming from above. The scintillator 
system consists of horizontal and vertical layers. The scintillator 
resolutions for muons are about $0.5\ ns$ in time and $11\ cm$ 
in position. 

Thanks to its large area, fine tracking granularity and electronics 
symmetry with respect to upgoing and downgoing flight directions, the 
MACRO detector is a proper tool for the study of
upward-travelling muons, generated by external $\nu$ interactions. 
Its mass ($\sim 5.3\ kton$) allows also to collect a statistically 
significant sample of neutrino events due to internal interactions. 

\section{Neutrino events in MACRO}
\begin{figure}[ht]
 \begin{center}
  \vspace{-0.65cm}
  \mbox{\epsfig{file=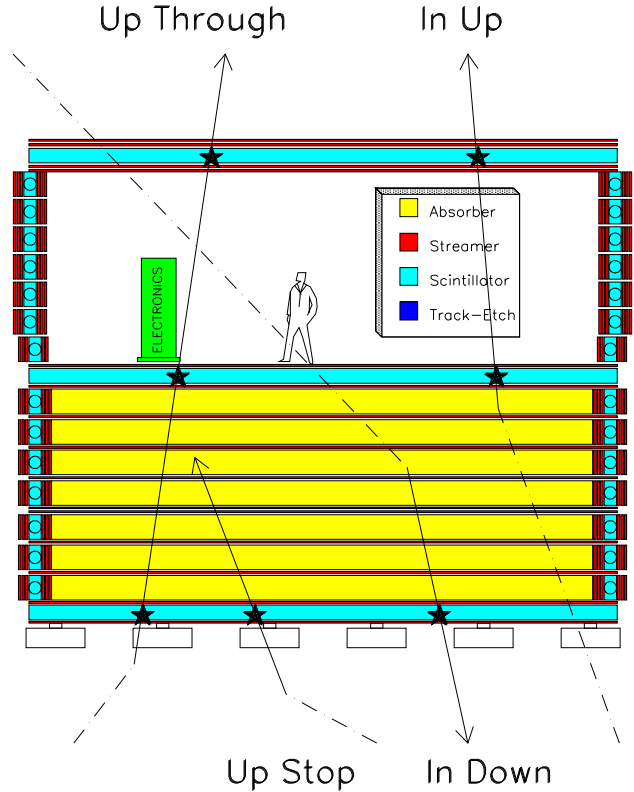,width=12cm}}
  \vspace{-1.0cm}
  \caption{\em {Cross section view of the detector and topology of the 
  neutrino induced events. The stars, the dashed lines and the full
  lines indicate scintillator hits, neutrino paths and charged 
  particle paths, respectively.\label{fig:topo}}}
  \vspace{-0.5cm}
 \end{center}
\end{figure}

\begin{figure}[ht]
 \begin{center}
  \vspace{-2cm}
  \mbox{\epsfig{file=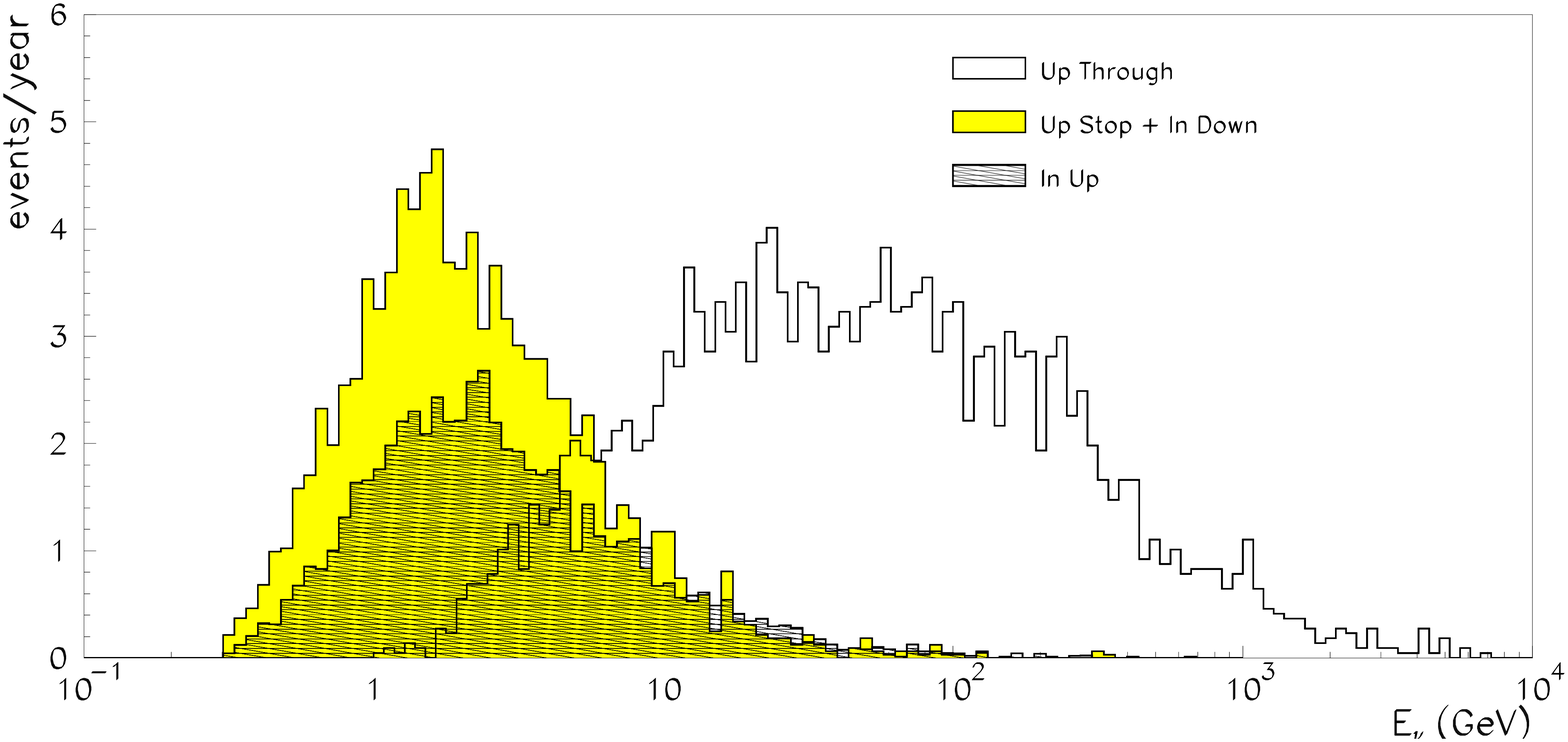,width=14cm}}
  \caption{\em {Distributions of the parent neutrino energy giving 
  rise to different kinds of events, estimated by means of Monte 
  Carlo simulation using the same cuts applied to the data. The 
  average energy is $\sim 100\ GeV$ for \lnf sample and 
  $\sim 4\ GeV$ for \le and \bo samples. 
  \label{fig:entopo}}}
  \vspace{-0.5cm}
 \end{center}
\end{figure}

Fig.~\ref{fig:topo} displays the different kinds of neutrino events here 
analyzed. Most of the detected particles are muons generated in $\nu_\mu$ 
Charged Current interactions. In Fig.~\ref{fig:entopo} the
parent neutrino energy distributions for the different event topologies 
are shown~:
\begin{itemize}
    \item \lnf - These tracks come from interactions in the rock 
        below MACRO and cross the whole detector ($E_\mu > 1\ GeV$). 
        The time information provided by scintillator counters 
        allows to know the flight direction by means of the 
        time-of-flight ({\it T.o.F. }) method. The data have been 
        collected in three periods, with different detector 
        configurations. In the first two periods (March 
        1989~--~November 1991, December 1992 -- June 1993) only 
        lower parts of MACRO were working\cite{ah95}. In the last 
        period (April 1994~--~February 1999) also the \atti was in 
        acquisition.
    \item \le - These partially contained events come from $\nu$ 
        interactions inside the apparatus. Also in this case the 
        analysis algorithm is based on the \tof calculation made
        possible by the \atti scintillator layers. Hence only the 
        data collected with the \atti (live-time $\sim 4.1$ years)
        are useful for this analysis. A not negligible percentage 
        ($13\ \%$) of events is expected to be induced by Neutral 
        Currents or $\nu_e$ CC interactions.
    \item \bo - This sample is composed by two subsamples :
        external interactions with upward-going track stopping in 
        MACRO ({\bf Up Stop}), interactions in the lower part of 
        the detector with a downgoing track ({\bf In Down}). 
        These events are recognized by means of topological criteria
        and the lack of time information makes difficult to 
        distinguish the two subsamples. Up to now we consider them
        as an unique sample. Assuming that neutrinos do not oscillate,
        it is expected that the number of \boa is almost equal to 
        the number of \bob and the contribution of Neutral Currents 
        and $\nu_e$ CC interactions is $\sim 10\ \%$. The \atti 
        is used also in the analysis of this sample as a veto
        for downgoing tracks. Therefore the analyzed data are those
        collected with the whole detector with the same effective 
        live-time of the \le sample.
\end{itemize}

\section{Analysis procedure}
The \tof method uses the formula
\be \label{beta}
\frac{1}{\beta} = \frac{c \times (T_1 - T_2)}{L}, \ee
where $T_1$ and $T_2$ are the times measured in lower and higher 
scintillator counters, respectively, and $L$ is the path between the 
two counters. Therefore $1/\beta$ results roughly $+1$ for 
downgoing tracks and $-1$ for upgoing tracks. In the sample of 
throughgoing muons it happens that a track hits 3 scintillator 
layers. In this case the time measurements are
redundant and $1/\beta$ is calculated by means of a linear fit
of the time as a function of the pathlength.

Several cuts are imposed to remove backgrounds from radioactivity 
and showering events which may cause failure in time reconstruction. 
Another background is connected to photonuclear processes such as
$\mu N \rightarrow \mu \pi X$ where low-energy upgoing particles are
produced at large angles by downgoing muons\cite{amb1}. The 
requirement of a minimum range of $200\ g/cm^2$ in the apparatus 
is applied to the \lnf sample in order to reduce drastically these 
low-energy tracks which mimic neutrino induced events when the 
downgoing muon misses MACRO.

\begin{figure}[thb]
 \begin{center}
%%  \vspace{-0.5cm}
  \vspace{-1.5cm}
  \mbox{\epsfig{file=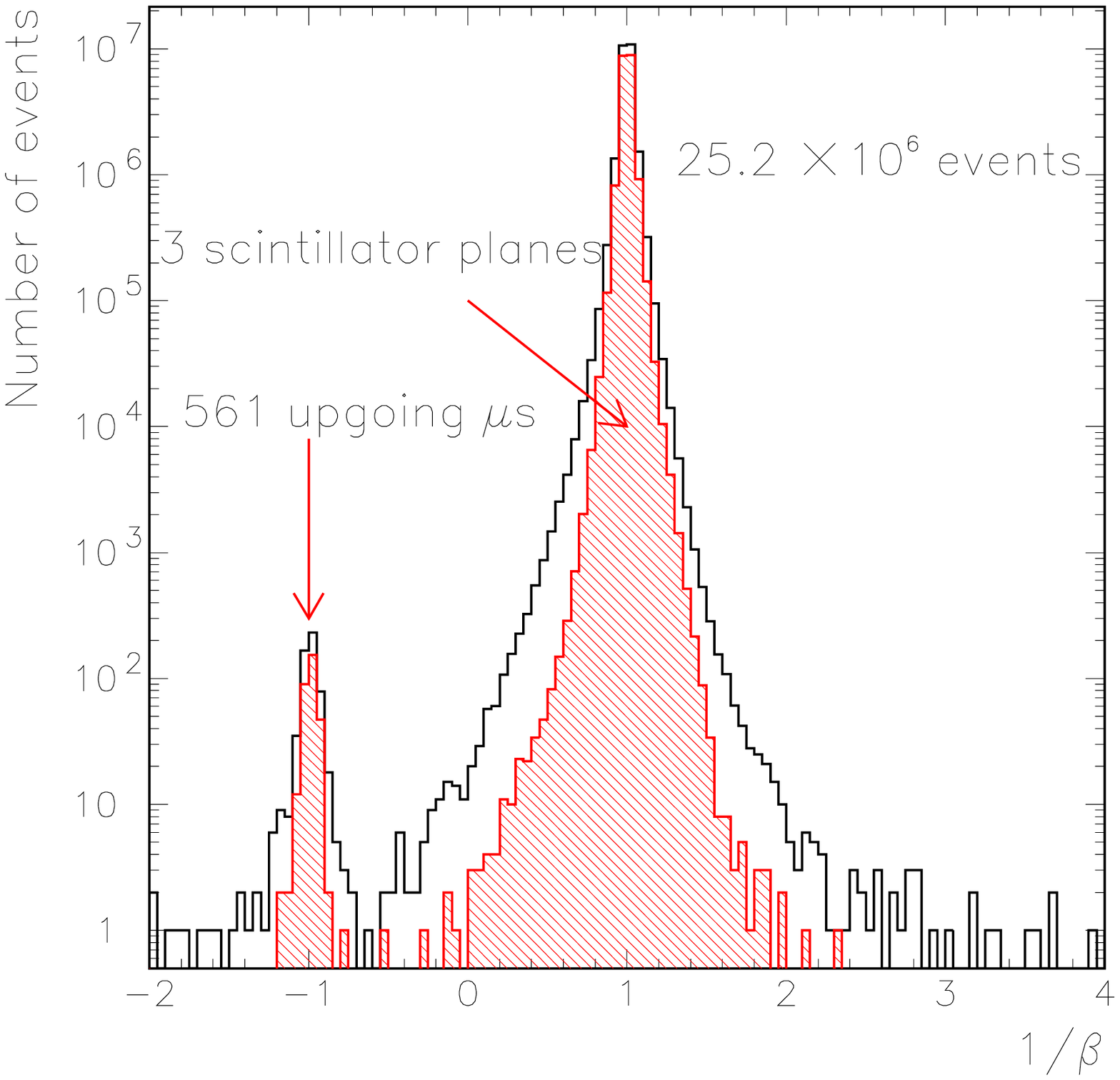,width=7cm}}
  \hspace{+0.5cm}
  \mbox{\epsfig{file=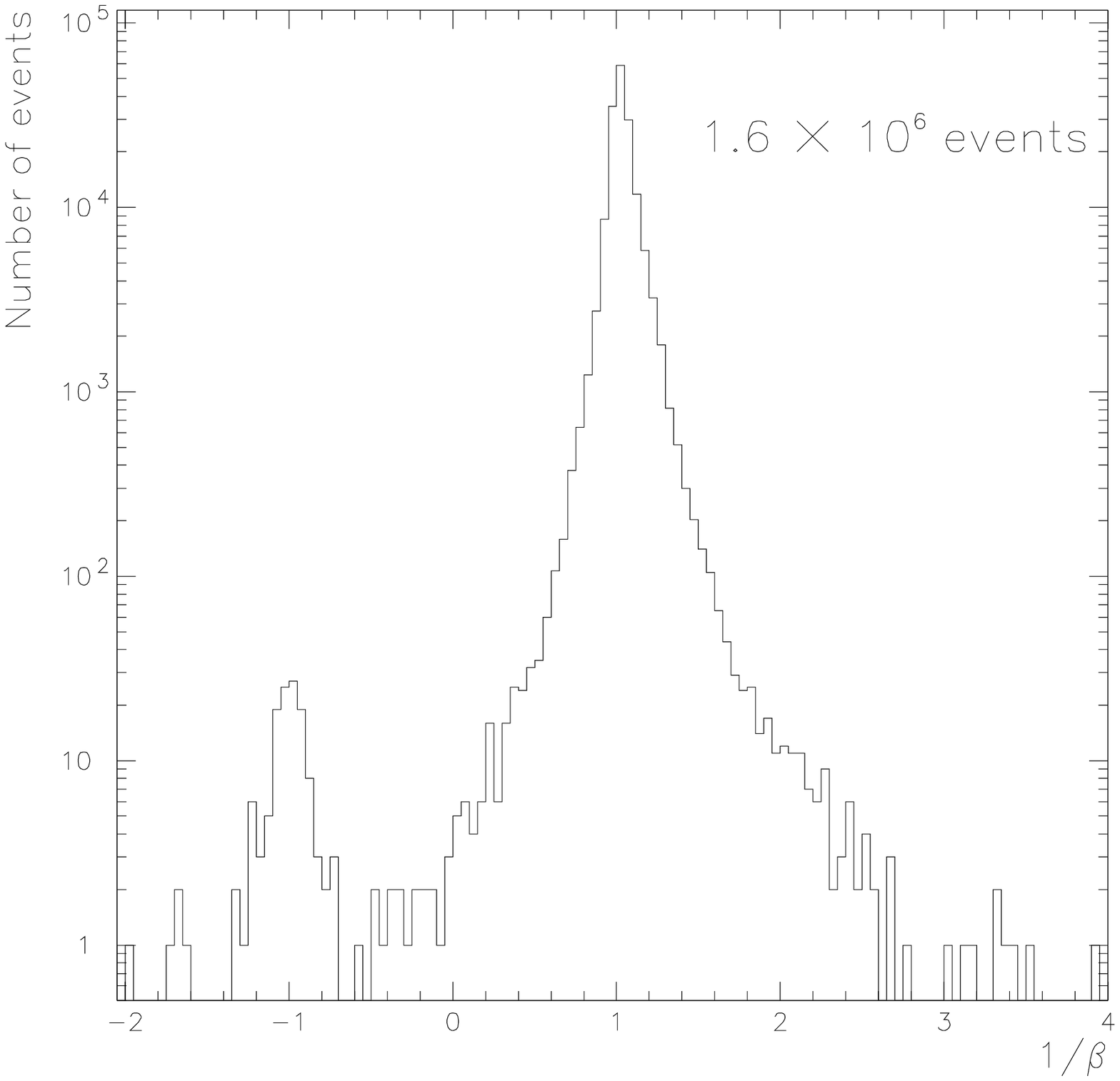,width=7cm}}
  \caption{\em {$1/\beta$ distributions after all analysis cuts
  for throughgoing (first plot) and partially contained (second 
  plot) tracks. The data are those collected when also the \atti 
  was in operation. The neutrino-induced signals are close to 
  $1/\beta~=~-1$. The shaded part in the first plot indicates 
  events whose \tof is calculated by means of a fit of three 
  times. In the second plot the events with $1/\beta~>~0$
  are downgoing atmospheric muons stopping in the detector.
  \label{fig:sbeta}}}
  \vspace{-0.5cm}
 \end{center}
\end{figure}

After all analysis cuts the signal peaks with $1/\beta \sim -1$ 
are well isolated for the first two samples (see Fig.~\ref{fig:sbeta}). 

The \bo events are identified via topological constraints. The main 
requirement is the presence of a reconstructed track crossing the 
bottom scintillator layer. All the track hits must be at least 
$1\ m$ far from the supermodule walls. The criteria used to verify 
that the event vertex (or $\mu$ stop point) is inside the detector 
are similar to those used for the \le search. The probability 
that an atmospheric muon produces a background event is negligible.
To reject ambiguous and/or wrongly tracked events which survive
automated analysis cuts, real events are directly scanned with the 
MACRO Event Display.
Because of the low energy of these events the minimum range 
of $200\ g/cm^2$ in the apparatus is not required. Therefore
the background due to upward going charged pions is estimated 
via simulation and subtracted on a statistical basis.

Expected rates and angular distributions have been estimated
assuming the atmospheric $\nu$ flux calculated by the Bartol
University group\cite{gais}\cite{agra}. The estimate of $\nu$
cross-section was based on GRV94\cite{grv} parton distribution
set, which varies by $+1\ \%$ the prediction with respect to the 
Morfin and Tung\cite{morf} parton distribution used in the past
by the MACRO Collaboration. For \le and
\bo samples also low-energy effects\cite{lipa} have been taken
into account. The propagation of muons in the rock was taken from
Lohmann {\it et al}\cite{lohm}. The uncertainty on the expected
muon flux is estimated $17\ \%$ for \lnf events and $25\ \%$ for
the other events. The apparatus and the data acquisition are fully
reproduced in a GEANT\cite{brun} based Monte Carlo program. Real
and simulated data are analyzed by means of the same procedure.
Particular care has been taken to minimize the systematic
uncertainty in the detector acceptance simulation. For the \lnf
sample different acceptance calculations, including separate
electronic and data acquisition systems, have been compared. For
each sample, two different analyses have been performed getting
the same results. Furthermore trigger and streamer tube
efficiency, background subtraction, effects of analysis cuts have
been in detail studied. The efficiency of the visual scanning
for the \bo sample has been estimated by analyzing real and 
simulated events after a random merging. The systematic error on 
the total number of events due to the acceptance has been estimated 
$6\ \%$ for \lnf sample. The uncertainty is higher ($10\ \%$) 
for low-energy samples because it depends strongly on data taking 
conditions, analysis algorithm efficiency and detector mass. 

\begin{figure}[thb]
 \begin{center}
  \mbox{\epsfig{file=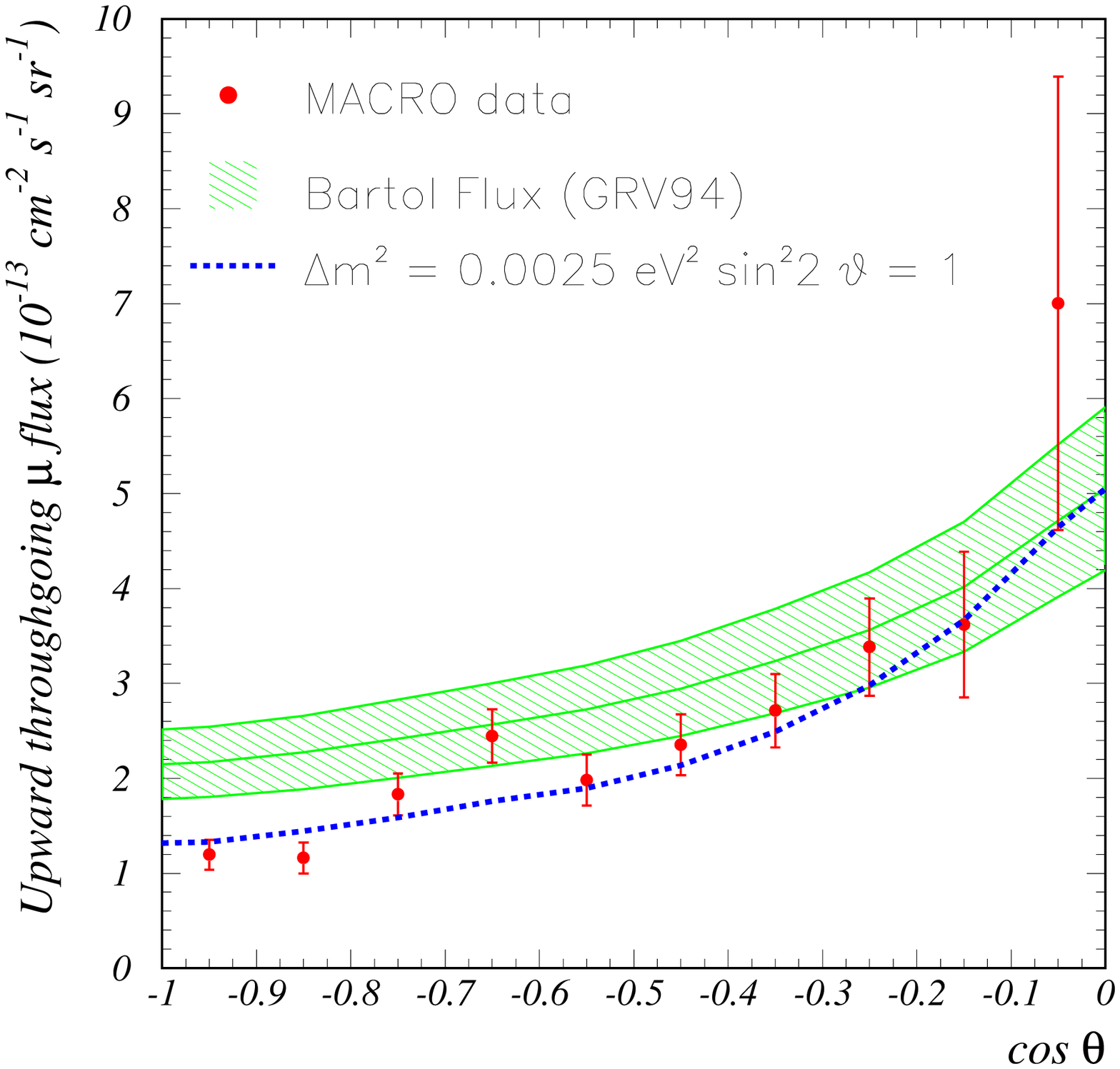,width=10cm}}
  \vspace{-0.4cm}
  \caption{\em {Measured and expected fluxes are shown for the \lnf 
  sample with a muon energy threshold of $1\ GeV$. The solid curve 
  and the shaded region show the expectation for no oscillations and 
  its uncertainty (the uncertainty on the predicted shape is much 
  smaller). The dashed line shows the predicition assuming 
  \mutau oscillation with maximal mixing and 
  $\Delta m^2 = 0.0025\ eV^2$.\label{fig:flux}}}
  \vspace{-0.5cm}
 \end{center}
\end{figure}

\begin{figure}[bht]
 \begin{center}
  \mbox{\epsfig{file=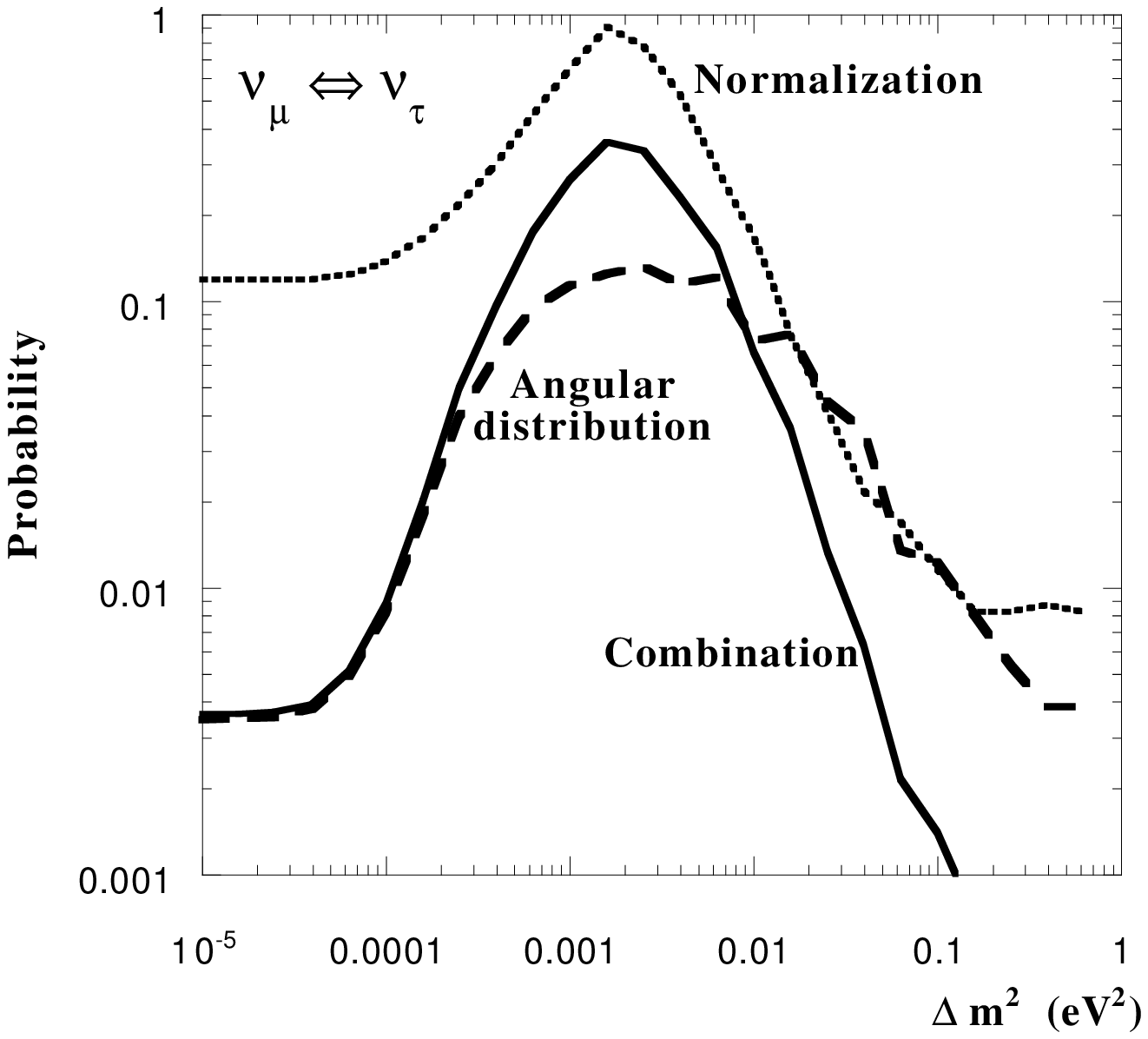,width=7cm}}
  \mbox{\epsfig{file=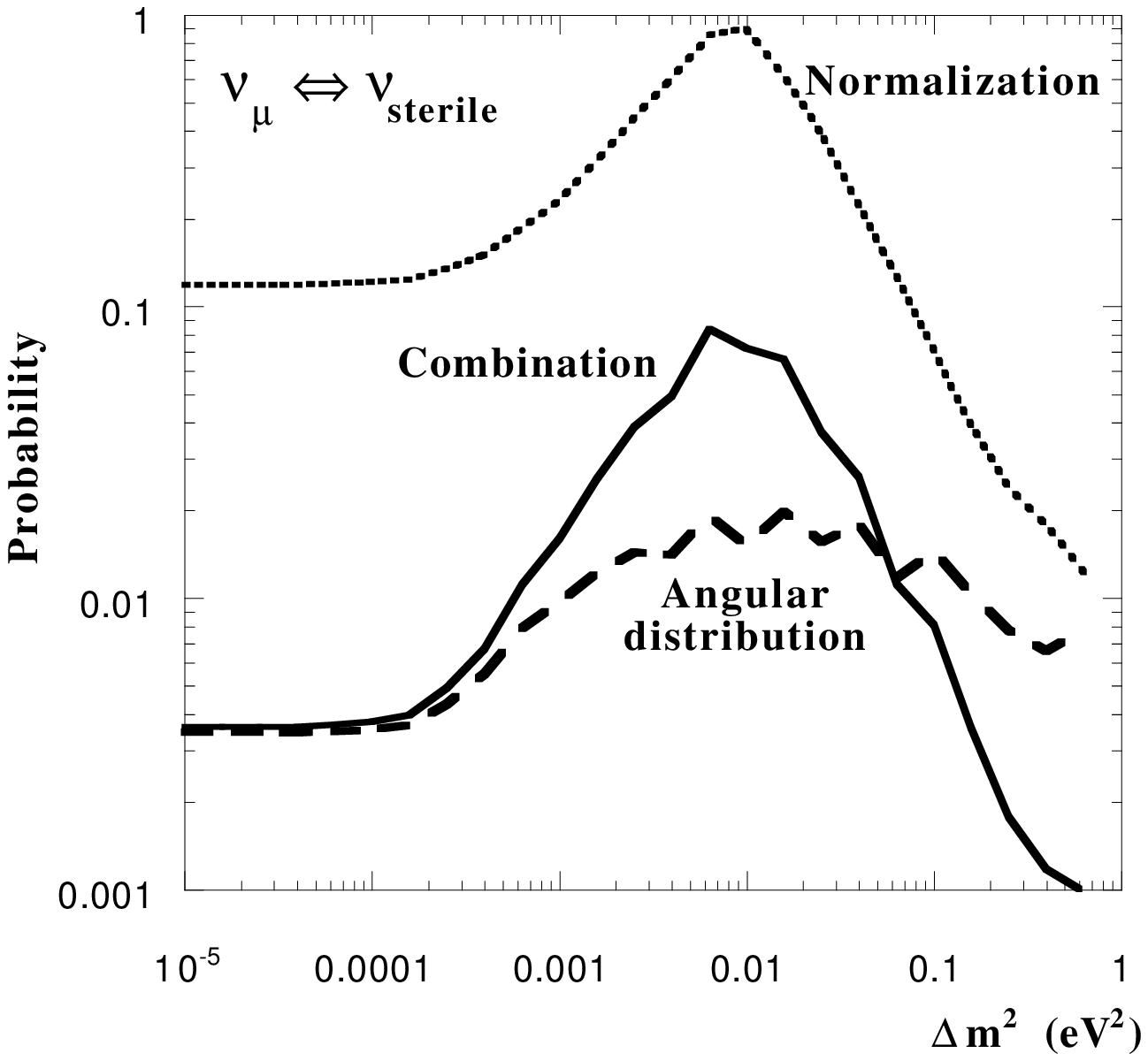,width=7cm}}
  \caption{\em {Probabilities of \lnf muon results assuming $\nu$ 
  oscillation with maximal mixing. The dotted line is referred 
  to the total number of events, the dashed line to the $\chi^2$ 
  of the normalized angular distribution and the continuos line to 
  their combination. The \mutau
  oscillation is assumed for the first plot, the 
  $\nu_\mu~\leftrightarrow~\nu_{sterile}$ for the second one.
  \label{fig:proba}}}
  \vspace{-0.5cm}
 \end{center}
\end{figure}

\section{High energy sample - Analysis results}
In the \lnf sample 642 events are in the signal range ($-1.25 <
1/\beta < -0.75$). Looking at the distribution of events outside
the signal peak, we estimate a background contamination of $12.5
\pm 6$ events. Furthermore $10.5 \pm 4$ events are expected to be
upgoing charged particles produced by downgoing muons. Finally we
expect that $12 \pm 4$ are not {\it true} \lnf events because they
are generated by neutrino interactions in the very bottom layer of
MACRO scintillators. Hence, removing these backgrounds, the
observed number of \lnf muons integrated over all zenith angles
results $607$. 

For this sample $825$ events are expected and the ratio of observed
to expected events is reported in Table~\ref{tab:nosci}.
In Fig.~\ref{fig:flux} the $\cos\theta$ distribution of the
measured flux is shown compared with the expected ones 
assuming stable or oscillating neutrinos. The data error bars 
are the statistical errors with an extension due to the 
systematic errors, added in quadrature. The observed zenith 
distribution does not fit well with the hypothesis of no oscillation, 
giving a maximum $\chi^2$ probability\footnote{The data are 
normalized to the predicition. The last bin near the horizontal 
is not taken into account because of the higher acceptance 
uncertainty and the background connected to scattering of 
quasi horizontal downgoing muons.} of only $0.35\ \%$. 
Combining normalization and angular shape the probability is
still very low ($0.36\ \%$). 

\begin{figure}[thb]
 \begin{center}
%%  \mbox{\epsfig{file=iso_plot.eps,width=7cm}}
  \mbox{\epsfig{file=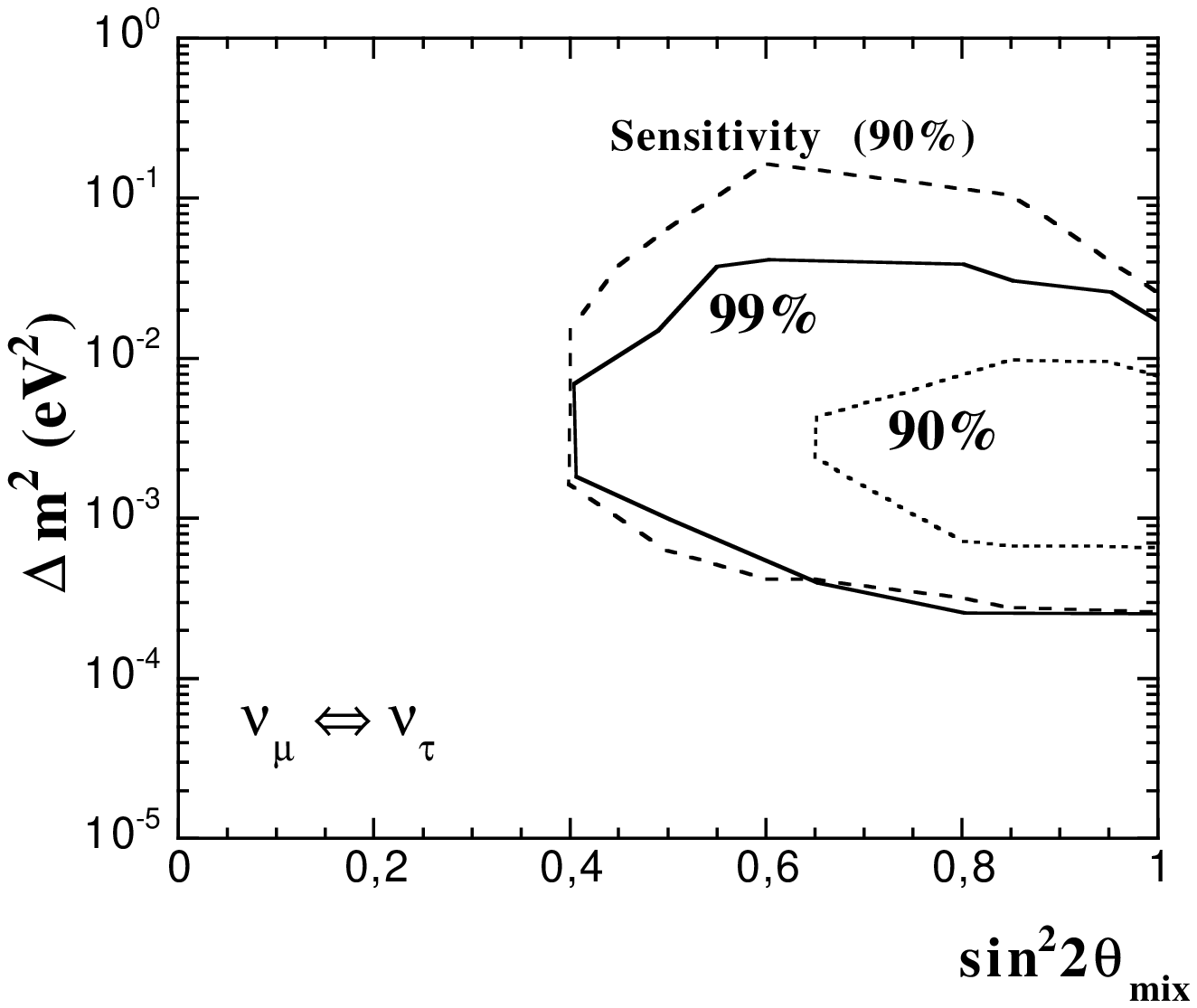,width=10cm}}
  \vspace{-0.4cm}
  \caption{\em {The results of the combined (normalization + 
  angular shape) analysis for \lnf sample are shown in terms of 
  confidence regions at the $90\ \%$ and $99\ \%$ levels 
  assuming $\nu_\mu~\leftrightarrow~\nu_\tau$. Since the best 
  probability is outside the physical region the confidence 
  interval regions are smaller than those expected from the 
  sensitivity of the experiment.
  \label{fig:exclu}}}
  \vspace{-0.5cm}
 \end{center}
\end{figure}

These results can be explained in the scenario of $\nu$ oscillation. 
Assuming \mutau the lower $\chi^2$ value for the angular distribution 
in the physical region\footnote{The best $\chi^2$ value is 10.6 in 
the unphysical range ($\sin^2 2\theta_{mix} = 1.5$).} is $12.5$ with 
maximal mixing and $\Delta m^2 \sim 0.0025\ eV^2$. In the first plot
of Fig.~\ref{fig:proba} the independent probabilities for normalization 
and angular shape and the combined probability are shown as functions 
of $\Delta m^2$ assuming maximal mixing. It is notable that 
total number of events and $\cos~\theta$ distribution indicate 
very close values of $\Delta m^2$. The second plot of 
Fig.~\ref{fig:proba} shows the same probabilities for 
$\nu_\mu \leftrightarrow \nu_{sterile}$ oscillation\cite{akh}\cite{liu}.
The maximum of the combined probability is $36.6\ \%$ for \mutau and 
$8.4\ \%$ for oscillation into sterile neutrino. 

Fig.~\ref{fig:exclu} shows the confidence regions at the $90\ \%$
and $99\ \%$ confidence levels based on application of the Monte Carlo
prescription by Feldman {\it et al}\cite{feld}. Also the sensitivity of
the experiment is plotted. The sensitivity is the $90\ \%$ contour 
which would result when the data are 
equal to the Monte Carlo prediction at the best-fit point. 

%
% le   116/202   fattore di riduzione 0.583 (mail di MS del 21 aprile)
% bo   193/274   fattore di riduzione 0.772 (mail di MS del 23 aprile)
%
\begin{table}[t]
 \begin{center}
  \begin{tabular}{|c||c|c|c|c||c|} \hline
Topology & Ratio & Statist. & Syst. & Theor. &    Ratio with     \\
         &       &  error   & error & error  & $\nu$ oscillation \\\hline
  \lnf   & 0.736 &  0.031   & 0.044 & 0.12   &        1.058      \\
  \le    & 0.57  &  0.05    & 0.06  & 0.14   &        0.98       \\
  \bo    & 0.71  &  0.05    & 0.07  & 0.18   &        0.92       \\\hline
  \end{tabular}
  \caption{\em {Ratios of observed on expected number of events
     for different event topologies. In the last column the ratio 
     is calculated assuming \mutau oscillation with the parameters 
     suggested by the \lnf angular shape ($\sin^2 2\theta_{mix} = 1$, 
     $\Delta m^2 = 0.0025\ eV^2$).\label{tab:nosci}}}
 \end{center} 
\end{table} 

\section{Low energy samples - Analysis results}
In the \le sample the uncorrelated background is estimated from 
the $1/\beta$ distribution. After the background subtraction $116$ 
events are accepted. The prediction is $202$ \le events.
In the \bo sample $193$ events survive the analyis cuts and the
visual scanning while $274$ events are expected.

\begin{figure}[thb]
 \begin{center}
  \mbox{\epsfig{file=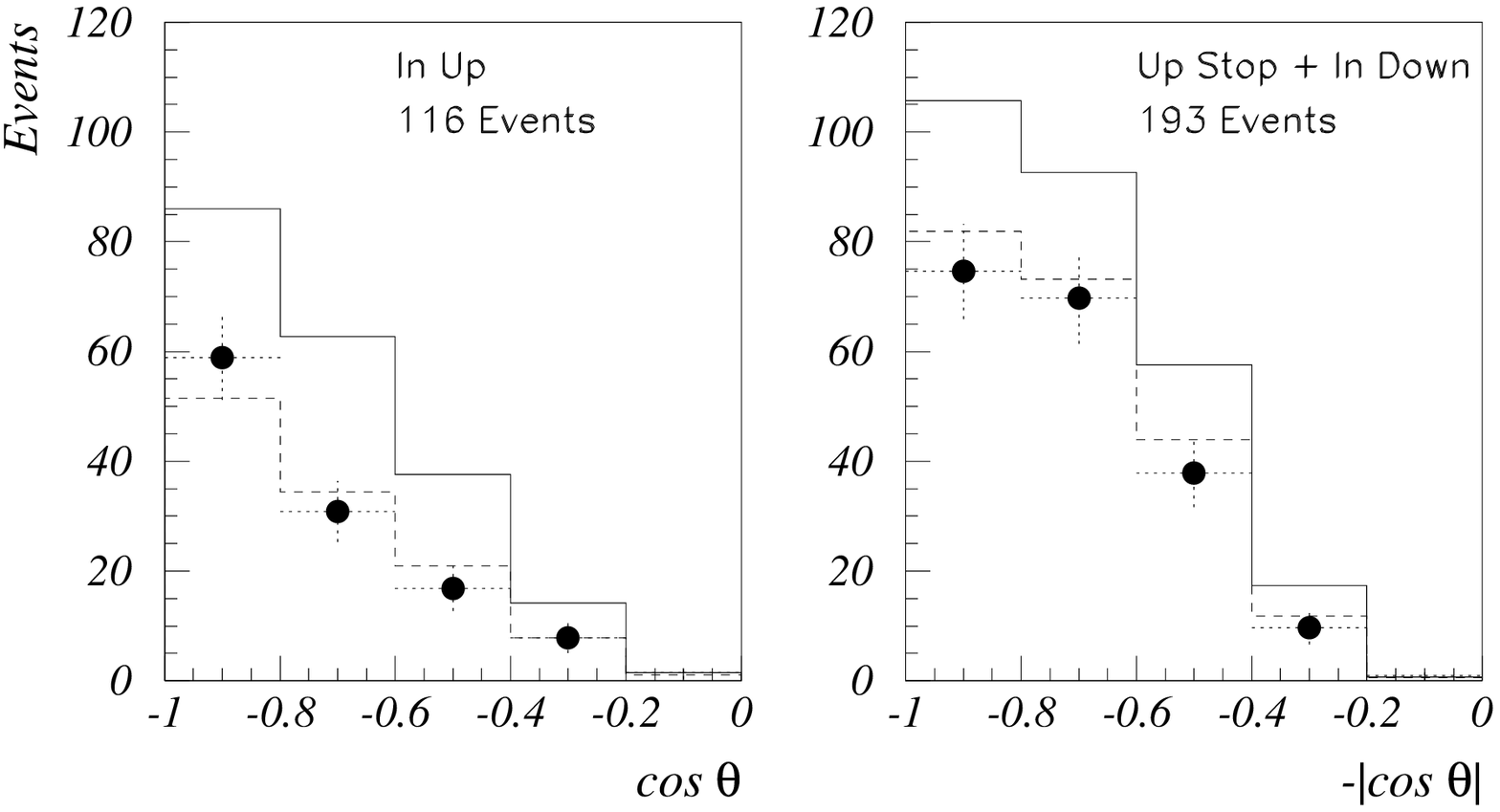,width=17cm}}
  \caption{\em {Comparison between measured and expected number of
  low-energy events versus $\cos~\theta$ (\le in the first plot, \bo
  in the second one). The dashed line is obtained assuming neutrino 
  oscillation with the parameters suggested by \lnf sample. In the
  second plot the absolute value of $\cos~\theta$ is used because
  the flight direction is unknown.
  \label{fig:doppia}}}
  \vspace{-0.5cm}
 \end{center}
\end{figure}

The ratios of the observed number of events to the prediction and the 
angular distributions of both samples are reported in 
Table~\ref{tab:nosci} and in Fig.~\ref{fig:doppia}. The low-energy 
$\nu_\mu$ samples show an uniform deficit of the measured number of 
events over the whole angular distribution with respect to the 
predictions based on the absence of neutrino oscillation. 

We note the agreement between the results for low-energy and \lnf 
events. Assuming the oscillation parameters suggested by higher energy 
sample, it is expected a $\sim 50\ \%$ disappearance of $\nu_\mu$ in \le 
and \boa samples because of the neutrino path (thousands of kilometres). 
No flux reduction is instead expected for \bob events whose neutrino 
path is of the order of tens of kilometres. The ratios and the angular 
distributions estimated assuming the $\nu$ oscillation are also 
reported in Table~\ref{tab:nosci} and in Fig.~\ref{fig:doppia}. 

\begin{figure}[ht]
 \begin{center}
  \mbox{\epsfig{file=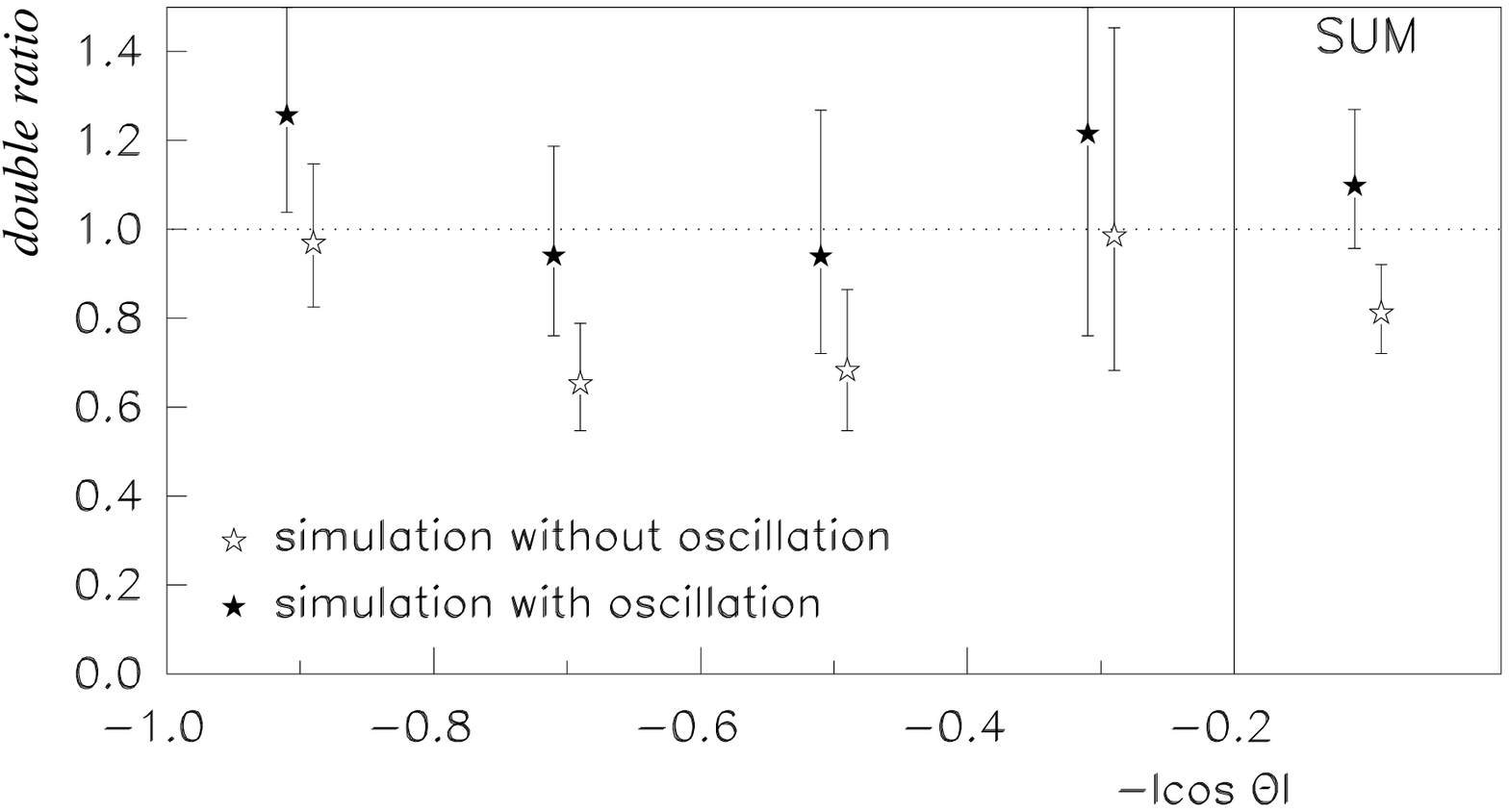,width=14cm}}
  \caption{\em {Double ratio (\ref{eq:RR}) in different hypotheses :
  stable $\nu_\mu$ and \mutau oscillation. The plotted errors
  are due to statistical, systematic and theoretical uncertainties.
  \label{fig:double}}}
  \vspace{-0.5cm}
 \end{center}
\end{figure}

In order to reduce the effects of uncertainties coming from neutrino
flux and cross section, the double ratio
\be \label{eq:RR}
\frac{(\frac{\le}{\bo})_{observed}}{(\frac{\le}{\bo})_{expected}} \ee
has been studied. A residual theoretical error ($5\ \%$) survives,
due to small differences between the energy spectra of the two samples.
Because of some cancellations the systematic uncertainty is also 
reduced to $\sim 6\ \%$. The value of the double ratio over the zenith 
angle distribution is shown in Fig.~\ref{fig:double}. Assuming the 
oscillation parameters ($\sin^2 2\theta_{mix} = 1$, 
$\Delta m^2 = 0.0025\ eV^2$) suggested by \lnf sample, the points are 
compatible with 1.

The probability to obtain a sum double ratio at least so far from 1 is 
$\sim 6 \%$ assuming no oscillation and the Bartol flux as the
%% true 
parent $\nu$ flux.

\section{Conclusions}

The flux and the shape of the zenith distribution for the \lnf 
sample favour \mutau oscillations. The experimental data have a 
$36.6\ \%$ probability assuming oscillation against $0.36\ \%$ 
assuming stable neutrino.

%% The probability of oscillation looking at the angular 
%% distributions only is $13\ \%$. The 
%% $\nu_\mu~\leftrightarrow~\nu_{sterile}$ model is slightly 
%% disfavored. 
%%
Therefore the new data confirm the MACRO published 
results\cite{ah95}\cite{amb2} with increased probability for the 
oscillation hypothesis.

The low-energy neutrino events are fewer than expected and
the deficit is quite uniform over the whole angular range. Also
these results suggest oscillation with maximal mixing and 
$\Delta m^2$ of a few times $10^{-3}\ eV^2$. 

The combined analysis of high and low energy data is in progress. 
Presently we stress the strong coherence of the MACRO results in 
different energy ranges and with different event topologies.

\end{document}